\documentclass[epj]{svjour}
\usepackage{times,graphicx,dcolumn}
\begin{document}
\title{Separated cross sections in \boldmath$\pi^0$ electroproduction
       at threshold at $Q^2=0.05\,\mathrm{GeV}^2/c^2$}
\titlerunning{Separated cross sections in $\pi^0$
  electroproduction at threshold at $Q^2=0.05\,\mathrm{GeV}^2/c^2$}
\author{
  M.~Weis		\inst{1}\and                
  P.~Bartsch		\inst{1}\and
  D.~Baumann		\inst{1}\and             
  J.~Bermuth		\inst{2}\and
  A.\,M.~Bernstein	\inst{3}\and
  K.~Bohinc		\inst{4}\and
  R.~B\"ohm		\inst{1}\and            
  M.~Ding		\inst{1}\and
  M.\,O.~Distler	\inst{1}\and         
  I.~Ewald		\inst{1}\and      
  J.\,M.~Friedrich	\inst{1}\fnmsep\thanks{{\it Present address:}
                        TU M\"unchen, Garching, Germany}\and
  J.~Friedrich		\inst{1}\and           
  M.~Kahrau		\inst{1}\and              
  M.~Kohl		\inst{5}\fnmsep\thanks{{\it Present address:}
                        M.I.T., Cambridge, USA}\and
  K.\,W.~Krygier	\inst{1}\and         
  A.~Liesenfeld		\inst{1}\and
  H.~Merkel		\inst{1}\and
  P.~Merle		\inst{1}\and               
  U.~M\"uller		\inst{1}\and  
  R.~Neuhausen		\inst{1}\and           
  M.\,M.~Pavan		\inst{3}\fnmsep\thanks{{\it Present address:}
                        TRIUMF, Vancouver, Canada}\and
  Th.~Pospischil	\inst{1}\and         
  M.~Potokar		\inst{4}\and    
  G.~Rosner		\inst{1}\fnmsep\thanks{{\it Present address:}
                        University of Glasgow, Glasgow, UK}\and
  H.~Schmieden		\inst{1}\fnmsep\thanks{{\it Present address:}
                        Universit\"at Bonn, Bonn, Germany}\and           
  M.~Seimetz		\inst{1}\fnmsep\footnotemark[5]\and
  S.~\v{S}irca		\inst{4}\fnmsep\inst{6}\and
  A.~Wagner		\inst{1}\and              
  Th.~Walcher		\inst{1}\\
[3mm]\textit{A1 Collaboration}
}
\mail{merkel@kph.uni-mainz.de}
\institute{
  Institut f\"ur Kernphysik,  Johannes Gutenberg-Universit\"at Mainz,
  D-55099 Mainz, Germany 
  \and
  Institut f\"ur Physik, Johannes Gutenberg-Universit\"at Mainz,
  D-55099 Mainz, Germany 
  \and
  Laboratory for Nuclear Science, Massachusetts Institute of Technology, 
  Cambridge, MA~02139, U.S.A.
  \and
  Jo\v{z}ef Stefan Institute, SI-1001 Ljubljana, Slovenia
  \and
  Institut f\"ur Kernphysik, Technische Universit\"at Darmstadt,
  D-64289 Darmstadt, Germany
  \and
  Dept. of Physics, University of Ljubljana, SI-1000 Ljubljana, Slovenia
}
\date{Received: 25 May 2007}
\abstract{ 
The differential cross sections $\sigma_0=\sigma_T+\epsilon \sigma_L$,
$\sigma_{LT}$, and $\sigma_{TT}$ of $\pi^0$ electroproduction from the
proton were measured from threshold up to an additional center of mass
energy of 40\,MeV, at a value of the photon four-momentum transfer of
$Q^2= 0.05\,\mathrm{GeV}^2/c^2$ and a center of mass angle of
$\theta=90^\circ$. By an additional out-of-plane measurement with
polarized electrons $\sigma_{LT'}$ was determined. This showed for the
first time the cusp effect above the $\pi^+$ threshold in the
imaginary part of the $s$-wave. The predictions of Heavy Baryon Chiral
Perturbation Theory are in disagreement with these data. On the other
hand, the data are somewhat better predicted by the MAID
phenomenological model and are in good agreement with the dynamical
model DMT.
}
\maketitle

\section{Introduction}                                        \label{sec:intro}

\sloppy

Over the last two decades, neutral pion photo- and electroproduction
at threshold have been established as a testing ground for Chiral
Perturbation Theory. First experimental attempts at Saclay
\cite{Mazzucato:1986dz} and Mainz \cite{beck90} to test the low energy
theorem for the $s$-wave multipoles at threshold
\cite{DeBaenst:1971hp,Vainshtein:1972ih} initiated the development of
an intense investigation of the threshold region with close
cooperation between experiment and theory.

Chiral Perturbation Theory (ChPT) is the effective field theory (low
energy approximation) of QCD
\cite{Weinberg:1978kz,Gasser:1983yg,Gasser:1983kx,Bernard:2006gx}. The
first reliable ChPT calculations of pion photoproduction in the direct
threshold region showed that the old low energy theorem has to be
modified, and that the $s$-wave prediction is a slow converging series
in this calculation \cite{BKM92}. For two $p$-wave combinations, new
low energy theorems were derived \cite{Bernard:1994gm}.

\begin{figure}
  \centerline{\includegraphics[width=0.7\columnwidth]{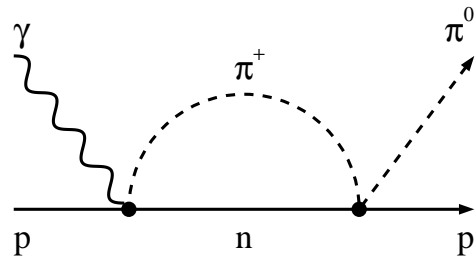}}
  \caption{The rescattering process responsible for the cusp in the
   $\pi^0$ \mbox{$s$-wave} amplitude at threshold.}
  \label{fig:rescattering}
\end{figure}

The energy dependence of the $s$-wave multipoles in the $\gamma
p\rightarrow \pi^0 p$ reaction has a smooth direct part and a rapidly
varying (cusp) part due to a $\pi^{+}$ rescattering contribution shown
in Fig.~\ref{fig:rescattering} \cite{Fal80,BKM95,Ber98}. Since the
$\pi^{+}$ production amplitude is dominated by the additional coupling
of the photon to the charge of the pion, this amplitude is much larger
than the production amplitude of neutral pions, leading to a sizable
unitarity cusp in the $\pi^{0}$ $s$-wave amplitude which appears in
the imaginary (real) part above (below) $n\pi^+$ threshold.

The predicted cusp effect in the real part of the transverse $s$-wave
amplitude $E_{0+}$, was established in photoproduction experiments at
SAL \cite{Berg96} and MAMI
\cite{Fuc96,Bernstein:1997vf,Schmidt:2001vg}. While these experiments
were only sensitive to the real part of this amplitude, one aim of
this work is to access the imaginary part of the $s$-wave
amplitudes. In electroproduction, this can be achieved by an
out-of-plane measurement with a longitudinally polarized electron beam
to access the interference cross section $\sigma_{LT'}$, which is
proportional to the imaginary part of the transverse and longitudinal
$s$-wave amplitudes $E_{0+}$ and $L_{0+}$, respectively.

The second aim of the present work is to improve the experimental
determination of the $p$-wave amplitudes. The existing
electroproduction experiments with large solid angle
\cite{Distler:1998ae,Merkel:2001qg} covered the energy range only up
to the first 4\,MeV above threshold. By an extension of the experiment
to an energy region of 40\,MeV above threshold, where the $p$-waves
clearly dominate, a more accurate separation of the cross sections is
possible.

\section{Notation and formalism}                          \label{sec:formalism}

In the one photon exchange approximation the electroproduction cross
section of pions can be written as (see \textit{e.g.} \cite{DT92})
\newlength\nmln\settoheight\nmln{$\sqrt{2_L(1)}$}
\begin{eqnarray}
  \frac{d\sigma(\theta,\phi)}{dE_e d\Omega_e d\Omega} &=& 
  \Gamma \left(\rule{0mm}{\nmln}
                \sigma_T(\theta) 
     + \epsilon~\sigma_L(\theta) 
     + \epsilon~\sigma_{TT}(\theta)\cos2\phi\right.\nonumber\\ 
  && + \sqrt{2\epsilon(1+\epsilon)}~\sigma_{LT}(\theta)\cos\phi\nonumber\\
  && + \left.h \sqrt{2\epsilon(1-\epsilon)}~\sigma_{LT'}(\theta)\sin\phi\right)
  \label{equ:cross}
\end{eqnarray}
with the virtual photon flux $\Gamma$, the transverse photon
polarization $\epsilon$, and the longitudinal beam polarization $h$.
In addition, the cross sections also depend on the center-of-mass
energy $W$ (or $\Delta W = W - W_{threshold}$) and on the photon
four-momentum transfer $q^2= - Q^2= \omega^2-\vec{q}^2$, with the
photon laboratory energy and momentum $\omega$ and $\vec{q}$.

\begin{figure}
\center
\includegraphics[width=\columnwidth]{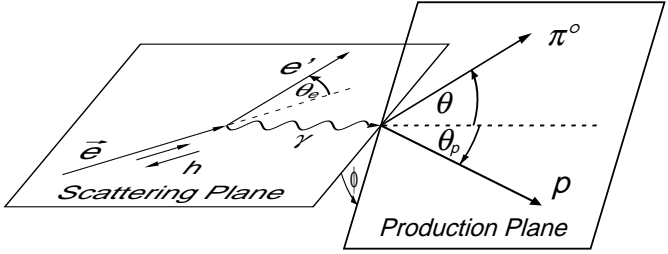}
\caption{Definition of the pion angles $\theta$ and $\phi$ in the
  laboratory frame.}
\label{fig:kinematics}
\end{figure}

The pion emission angles $\theta$ and $\phi$ are defined relative to
the direction of the momentum transfer
$\vec{q}$. Figure~\ref{fig:kinematics} shows the definition of the
angles in the laboratory frame. In the following, these angles are
used in the photon-proton c.m. system.

In this experiment, the photon polarization $\epsilon$ was not varied,
so the differential cross section
\begin{equation}
  \sigma_0(\theta) = \sigma_T(\theta) + \epsilon \sigma_L(\theta)
\end{equation}
could not be separated further.

Since $\sigma_{LT'}$ is small compared to the unpolarized cross
section, it is useful to define the asymmetry $A_{LT'}(\theta)$ to
minimize systematic errors:
\begin{equation}
  A_{LT'}(\theta) = \frac{\sigma^+ - \sigma^-}{\sigma^+ + \sigma^-} =
    \frac{\sqrt{2\epsilon(1-\epsilon)}\,\sigma_{LT'}(\theta)}
    {\sigma_T(\theta) +\epsilon\,\sigma_L(\theta)
    -\epsilon\,\sigma_{TT}(\theta)}
\label{eqn:asym}
\end{equation}
where $\sigma^+$ and $\sigma^-$ are the differential cross sections
for $\phi=90^\circ$ with beam polarization parallel and antiparallel
to the beam direction, respectively.

In the threshold region, all multipole amplitudes with angular
momentum $l\ge 1$ are negligible. The cross sections can be further
decomposed in $s$- and $p$-wave multipoles as
\begin{eqnarray}
  \sigma_{T}(\theta) & = & 
    p/{k_\gamma}\left(A + B \cos\theta + C\cos^2\theta \right), \nonumber \\
  \sigma_{L}(\theta) & = &
    p/{k_\gamma}\left(A' + B'\cos\theta +C' \cos^2\theta \right), \nonumber \\
  \sigma_{TT}(\theta)  & = &
    p/{k_\gamma}\left(F \sin^2\theta\right),\nonumber\\
  \sigma_{LT}(\theta) & = &
    p/{k_\gamma}\left(D \sin\theta + E \sin\theta\cos\theta\right),\nonumber\\
  \sigma_{LT'}(\theta) & = &
    p/{k_\gamma}\left(G \sin\theta + H \sin\theta\cos\theta\right),
 \label{equ:multi}
\end{eqnarray}

where $p$ is the pion center of mass momentum and $k_\gamma = (W^2 -
m_p^2 + q^2)/ (2 W)$ is the equivalent real photon energy. The angular
coefficients contain two $s$-wave and five $p$-wave multipole
combinations \cite{DT92,BKM96a}. By measuring the helicity asymmetry
with an out of plane measurement, we obtain $\sigma_{LT'}$ which is
proportional to the imaginary part of $(E_{0+}^* P_5 +
L_{0+}P_2^{*})$.

In this energy region the $p$-wave multipoles are approximately real
so we are directly sensitive to the imaginary part of the $s$-wave
multipoles.

\section{Experiment}                                     \label{sec:experiment}

The experiment took place at the three-spectrometer setup of the A1
Collaboration at MAMI (see \cite{Blo98} for a detailed
description). It was performed and analyzed based on the same
techniques described in
refs.~\cite{Distler:1998ae,Merkel:2001qg}. From Eq.~\ref{equ:cross}
one can see that for a separation of the cross sections, a measurement
at $\phi=0^\circ, 90^\circ, 180^\circ$ is the best choice.  A
measurement at $\phi=90^\circ$ requires out-of-plane detection.
				   
Spectrometer B of the A1 setup can be moved out of plane up to a
cartesian angle of $10^{\circ}$ in the laboratory frame
\cite{weism}. By choosing spectrometer B for the electron detection at
a small forward angle, we were able to cover $\phi=90^\circ$ up to a
center-of-mass momentum of $100\,\mathrm{MeV/c}$ within three
kinematical settings. In addition, four in-plane settings were chosen
to separate the cross sections for $\phi=0^\circ$ and
$\phi=180^\circ$.

\def\deg{$^\circ$}
\def\MeVc{$\left(\frac{\mathrm{MeV}}c\right)$}
\begin{table}
  \caption{Kinematical settings for the central trajectory of the
    spectrometers.  The symbols $\theta_e$ and $\phi_e$ mark the
    in-plane and out-of-plane cartesian angles of spectrometer
    B. Common for all settings are $Q^2=0.05\,\mathrm{GeV}^2/c^2$,
    $\theta=90^\circ$, $\epsilon=0.933$,
    $E_{beam}=854.49\,\mathrm{MeV}$.}
  \label{tab:settings}
  \begin{tabular*}{\columnwidth}{%
    @{\extracolsep{\fill}}D{-}{-}{3}@{~}rc@{~}cc@{~}c@{~}c}
    \hline\noalign{\smallskip}
    \multicolumn{2}{c}{pion CMS} & \multicolumn{2}{c}{Spectrometer A} 
    & \multicolumn{3}{c}{Spectrometer B}\\
    \multicolumn{1}{c}{$p$}&$\phi$&$\theta_p$&$p_p$&$\theta_{e}$&
    $\phi_{e}$&$p_{e}$\\[5pt]
   \multicolumn{1}{c}{\MeVc}&&&\MeVc&&&\MeVc\\[5pt]
   \noalign{\smallskip}
   \hline
   \noalign{\smallskip}
    0-50  &       & 44.6\deg & 230 & 16.8\deg &    0\deg & 652 \\
   45-85  &  0\deg& 52.4\deg & 243 & 16.9\deg &    0\deg & 662 \\
   65-100 &  0\deg& 58.1\deg & 263 & 17.2\deg &    0\deg & 662 \\
   45-85  & 90\deg& 43.4\deg & 243 & 16.5\deg & 3.96\deg & 652 \\
   65-100 & 90\deg& 40.1\deg & 263 & 15.4\deg & 7.65\deg & 652 \\
   45-85  &180\deg& 34.4\deg & 241 & 16.9\deg &    0\deg & 662 \\
   65-100 &180\deg& 22.1\deg & 259 & 17.2\deg &    0\deg & 662 \\
   \noalign{\smallskip}
   \hline
  \end{tabular*}
\end{table}

Table \ref{tab:settings} summarizes the kinematical settings including
the approximative acceptance range in the center-of-mass momentum. For
all settings, the four-momentum transfer of the photon was fixed to
the value of $Q^2 = 0.05\,\mathrm{GeV^2}/c^2$ used in former
experiments \cite{Merkel:2001qg}. At the beam energy of $E_0 =
854.49\,\mathrm{MeV}$, adjusted to optimize the spin precession angle
between polarized beam source and spectrometer setup, the photon
polarization was $\epsilon=0.933$.

The polarized electron source \cite{pole} of MAMI was run at a
polarization of 75\% at an average current of $5\,\mathrm{\mu A}$. The
beam polarization was determined by an elastic $H(\vec{e},e'\vec{p})$
measurement with a focal plane polarimeter in spectrometer A to an
accuracy of $\pm 5\%$ absolute. To avoid systematic errors, the beam
helicity was switched stochastically with a mean frequency of
$1\,\mathrm{Hz}$.

A subcooled liquid Hydrogen target with a oblong shape and a length of
5\,cm was used. The beam was moved across the target in a controlled
way by a fast raster magnet to avoid boiling of the Hydrogen. The beam
current of 5\,$\mu$A corresponds to a luminosity of $L=6.7\,
\mathrm{MHz/\mu barn}$. A total effective beam time of 400\,h was
accumulated.

Spectrometer A with a solid angle acceptance of 21\,msr was used for
the detection of the recoil proton, and spectrometer B with an
acceptance of 5.6\,msr was used for the detection of the scattered
electron. The focal planes of both spectrometers were equipped with
four layers of vertical drift chambers for position and angular
resolution and two layers of scintillators for trigger and timing
resolution. In addition, spectrometer B was equipped with a gas
Cerenkov detector for the suppression of charged pions. An overall
momentum resolution of $\delta p/p < 10^{-4}$ and an angular
resolution of better than 3\,mrad was achieved. After correction for
the flight path in the spectrometers, a coincidence time resolution of
2\,ns FWHM was determined.
 
\section{Analysis and error estimation}                    \label{sec:analysis}

\begin{figure}
  (a)\\\includegraphics[width=\columnwidth]{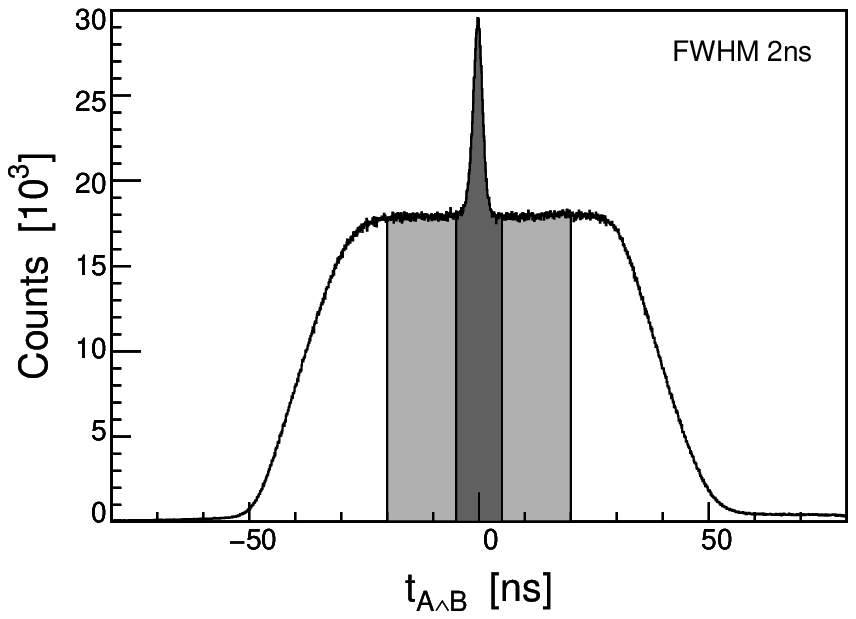}\\
  (b)\\\includegraphics[width=\columnwidth]{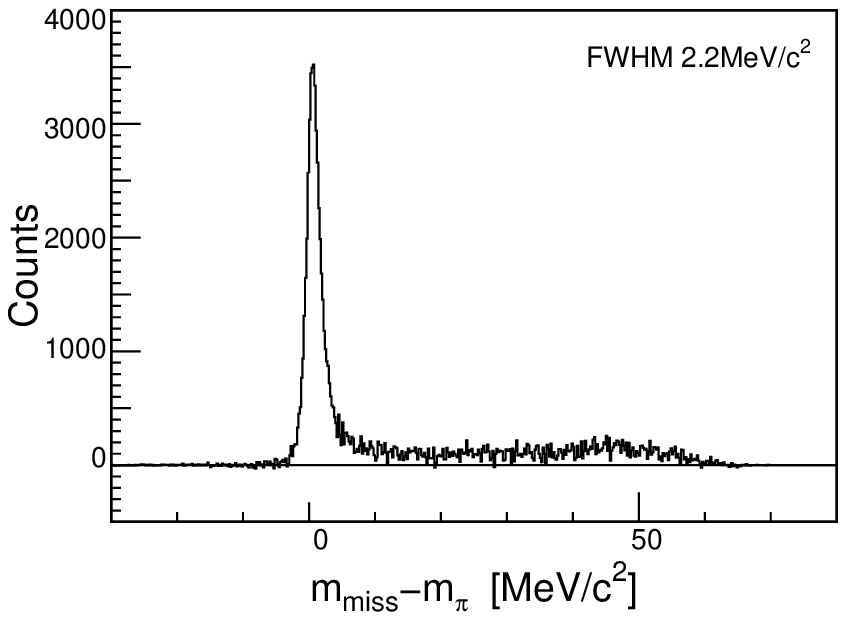}
  \caption{Reaction identification by coincidence time (a) and
    background subtracted missing mass (b). The events in the dark
    shaded area were defined as true coincidences, while the events in
    the light shaded area were used to subtract the background of
    random coincidences.}
  \label{TimeMM}
\end{figure}

\begin{figure*}
  \parbox{\columnwidth}{\center (a) Model dependent\\[4mm]
    \includegraphics[width=\columnwidth]{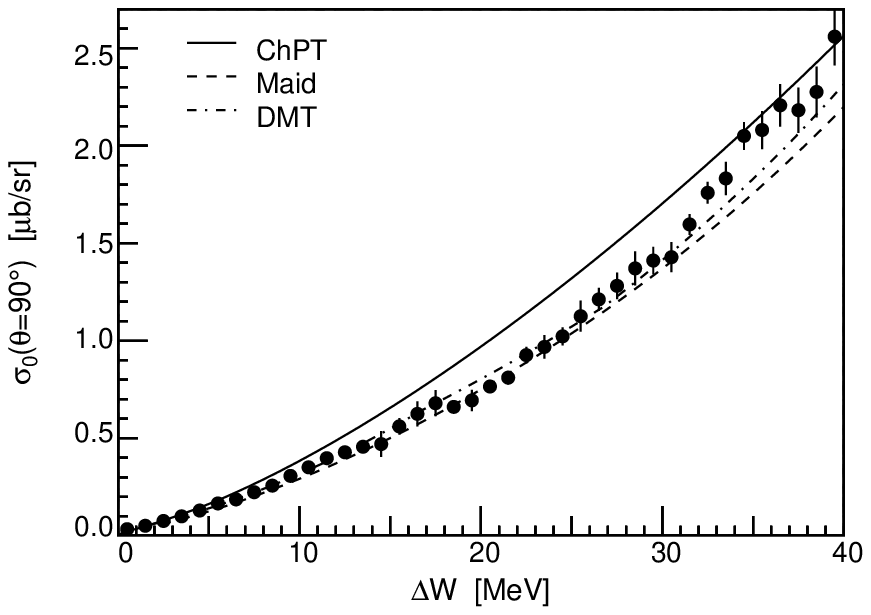}\\[5mm]
    \includegraphics[width=\columnwidth]{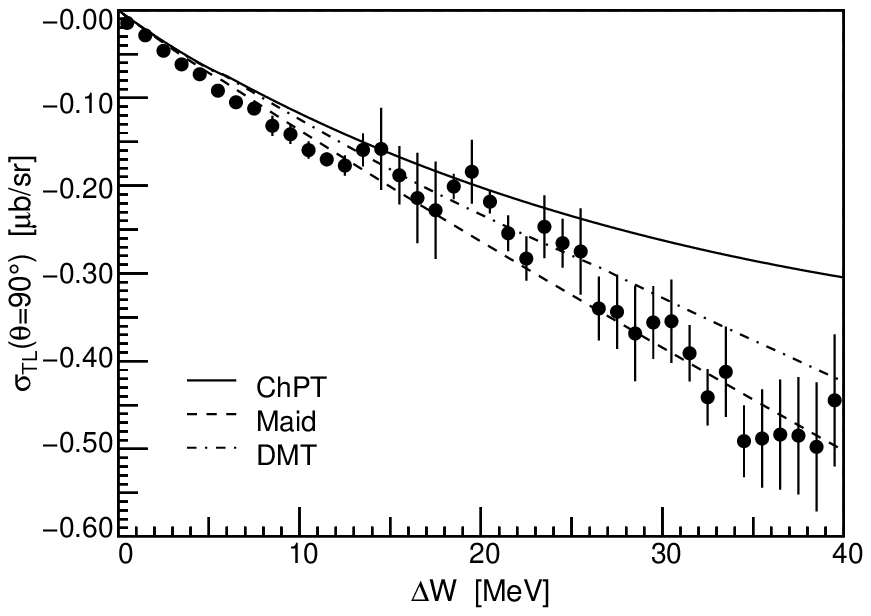}\\[5mm]
    \includegraphics[width=\columnwidth]{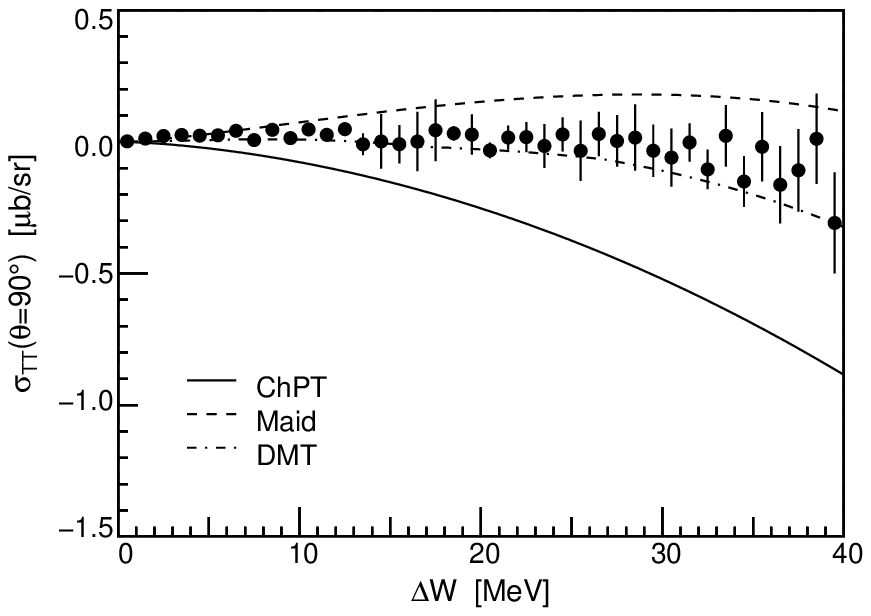}
  }
  \hfill
  \parbox{\columnwidth}{\center (b) Model independent\\[4mm]
    \includegraphics[width=\columnwidth]{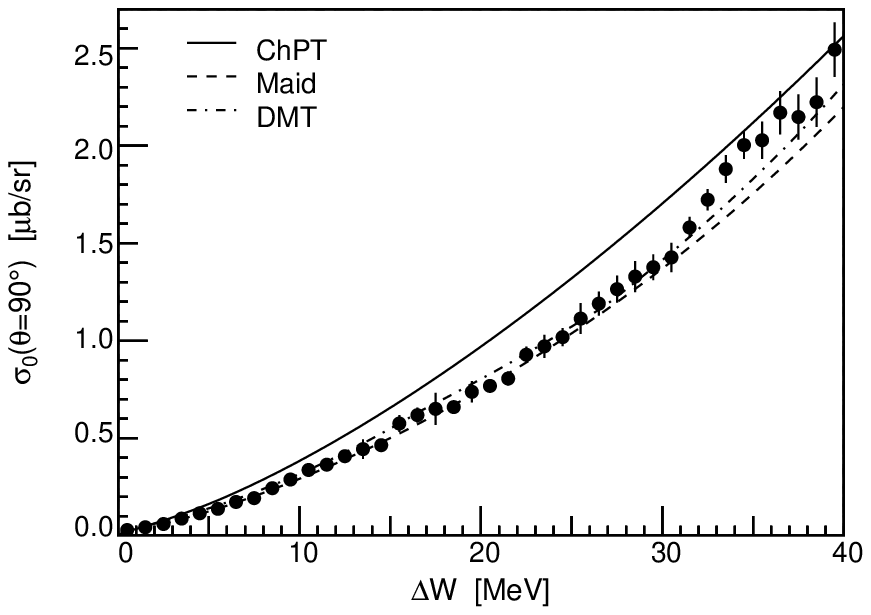}\\[5mm]
    \includegraphics[width=\columnwidth]{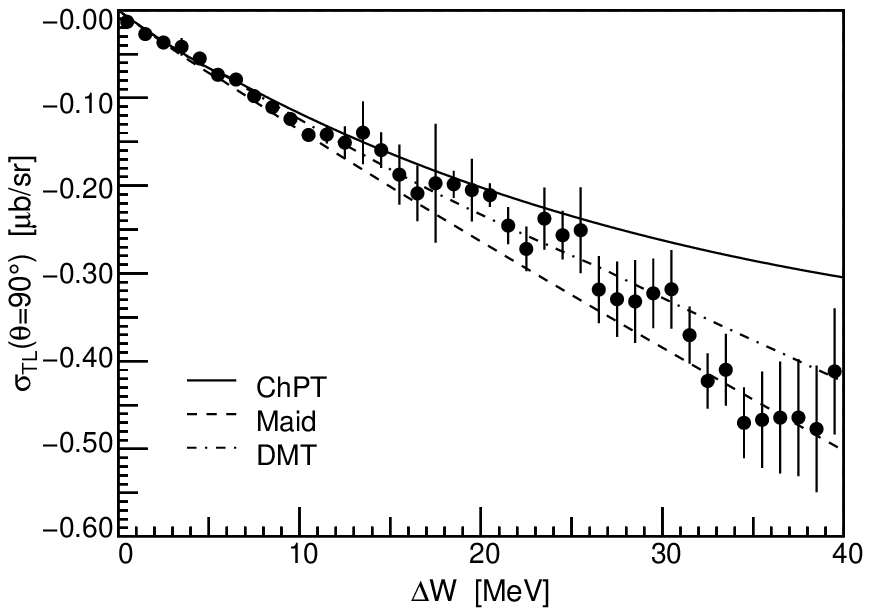}\\[5mm]
    \includegraphics[width=\columnwidth]{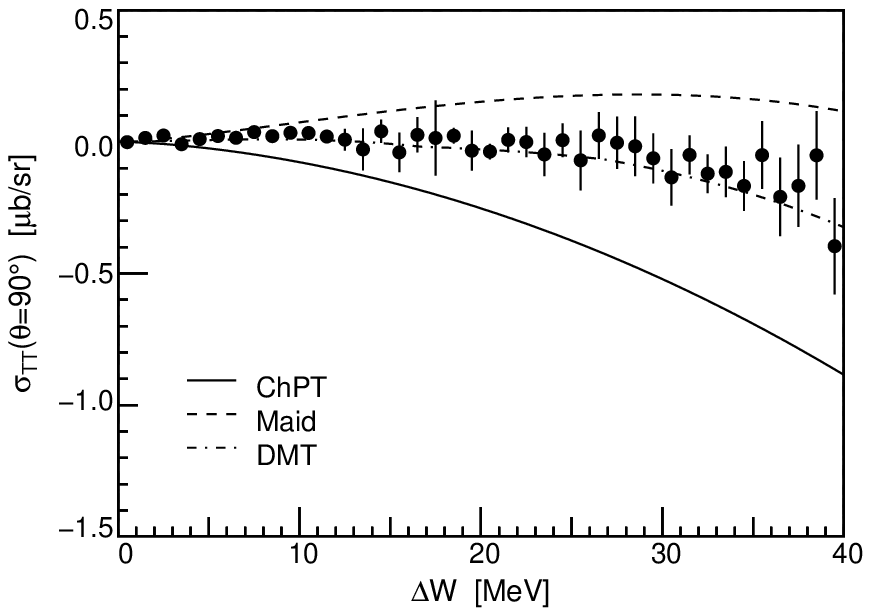}
  }
  \caption{The separated cross sections $\sigma_0$, $\sigma_{LT}$, and
    $\sigma_{TT}$ at $\theta=90^\circ$. Column (a) shows the result
    with model dependent analysis using the full statistics, while
    column (b) shows the model independent analysis with narrow
    kinematical cuts. The solid line shows the calculation in HBChPT
    \protect\cite{BKM96a}, the dashed line shows the phenomenological
    model MAID (2003 fit) \cite{DHKT99}, and the dash-dotted line
    shows the DMT model\cite{Kamalov:2001qg}.}
  \label{fig:structurefunctions}
\end{figure*}

In a first step, the electron and proton were identified by the time
of flight. The coincidence time was corrected by the flight path
length inside the spectrometers of about 10\,m. Figure~\ref{TimeMM}
(a) shows the corrected coincidence time distribution. The central
peak of this distribution was used as the yield of true coincidences,
while the events in the side bands were used to estimate the
background of random coincidences.

The reaction $H(e,e'p)\pi^0$ was identified by a cut on the missing
mass given by the four-momentum balance $m_{miss}^2 =
(e_{in}+p_{in}-e_{out}-p_{out})^{2}$. Figure~\ref{TimeMM} (b) shows
the missing-mass distribution after subtraction of the background of
random coincidences. The pion peak shows a missing-mass resolution of
$2.2\,\mathrm{MeV}/c^2$ and a radiative tail to higher masses. The
observed increase in radiative events at high masses was caused by the
increase in acceptance and phase space.

The integral of the accepted phase space was calculated with the Monte
Carlo method. In this calculation, all resolution and efficiency
effects and radiative effects according to the formulas of
ref.~\cite{MoTsai69} were included. The differential cross section
$\sigma(\theta,\phi)$ was extracted for several bins in $W$, $\theta$,
and $\phi$. Since the acceptance in $Q^2$ and $\epsilon$ was narrow,
no binning in these variables was required.

To separate the cross sections a $\chi^2$-fit of the $\phi$-dependence
as given by Eq.~\ref{equ:cross} was performed for each bin in the
center of mass energy $W$ and with a cut on the pion angle of
$75^\circ < \theta< 105^\circ$ .

By the cut in $\theta$ a large fraction of events was lost. In
addition, the acceptance in $\theta$ was different for each
kinematical setting. To overcome these two problems, a second, model
dependent method was used to separate the cross sections. In this
method, the phenomenological model MAID \cite{DHKT99} was used as
parameterization of the cross section.  For each event, the
differential cross section was projected to the nominal kinematics at
$\theta=90^\circ$ by
\begin{equation}
  \sigma(90^\circ,\phi) = \sigma(\theta,\phi) 
  \frac{MAID(90^\circ,\phi)}{MAID(\theta,\phi)}.
\end{equation}
By this method, the statistical error could be reduced for the price
of an additional model error.

The asymmetry $A_{LT'}$ was extracted by the helicity asymmetry
defined in Eq.~\ref{eqn:asym}, including an additional background term
in the denominator. The per-event-helicity and the extended $\phi$
acceptance were taken into account by a weight factor $h~\sin\phi$ for
each event.

The systematic errors in detection efficiency, luminosity calculation,
and reaction identification could be neglected compared to the
statistical error. The only notable systematic error was assigned to
the first three energy bins. Directly at threshold, the extracted
cross section is very sensitive to the absolute momentum calibration
of the electron arm, which was calibrated to 300\,keV$/c$ by a
measurement of elastic electron-proton scattering. For the first three
bins, this results in systematic errors of 43\%, 21\%, and 2\%
respectively. For $\Delta W > 3\,\mathrm{MeV}$, the corresponding
error is below 1\%. For all other systematic effects an overall
systematic error of 3\% was determined \cite{Blo98}.

The model error of the model dependent separation of the cross
sections was estimated by varying the multipoles of the model up to
$l=1$ separately by $\pm 10\%$. This results in a model error in the
extracted cross section of less than 5\%.  For the helicity asymmetry
$A_{LT'}$ the dominant systematic error was caused by the measurement
of the beam helicity $h$, which was known to 5\%. This error is
negligible compared to the statistical error.

\section{Results}                                           \label{sec:results}

Figure.~\ref{fig:structurefunctions} shows the separated unpolarized
cross sections $\sigma_0$, $\sigma_{LT}$, and $\sigma_{TT}$. Column
(a) shows the model dependent analysis, column (b) shows the model
independent analysis. Both results are clearly consistent within the
error bars. The statistical advantage of the better utilization of the
full data sample is compensated by the large systematic model
error. Of course, the size of this error is disputable: the choice of
a variation of 10\% in the leading multipoles is somewhat arbitrary
and the authors are convinced, that this choice is conservative. In
the future, better knowledge of all multipoles can clearly improve
this analysis.

Since the model dependent analysis shows that the model independent
analysis is not affected by the acceptance matching, we will discuss
only the model independent analysis. The data points of this analysis
are compiled in Table~\ref{tab:structurefunctions}.

The cross sections are compared with three theoretical
calculations. The solid line in Fig.~\ref{fig:structurefunctions}
shows a calculation in Heavy Baryon Perturbation Theory by V. Bernard
\textit{et al.} \cite{BKM96a}. As already noticed in
ref.~\cite{Merkel:2001qg}, this calculation overestimates the cross
section at $Q^{2} = 0.05 \,\mathrm{GeV}^{2}/c^2$. The reason for this
might be that some of the low energy parameters of this calculation
were fitted to data at a photon virtuality of
$Q^2=0.1\,\mathrm{GeV}^2/c^2$. This value may be somewhat too large
for the validity of chiral perturbation theory. Therefore, a refit of
the low energy parameters to low $Q^2$ data might resolve some of this
discrepancy. This would still leave the issue of the $Q^2$
dependence~\cite{Merkel:2001qg} as an unsolved problem. The more
likely solution might be to carry out the ChPT calculation to higher
order. The presently published calculation was carried out to order
${\cal{O}}(p^3)$ for the $p$-waves with some extra counter terms of
higher order for the $s$-waves.

The two phenomenological models MAID \cite{DHKT99} and DMT
\cite{Kamalov:2001qg} both agree well with the unpolarized cross
sections $\sigma_{0}$ and $\sigma_{LT}$. The small interference cross
section $\sigma_{TT}$, however, is described only by the DMT model.

\begin{figure}
  \includegraphics[width=\columnwidth]{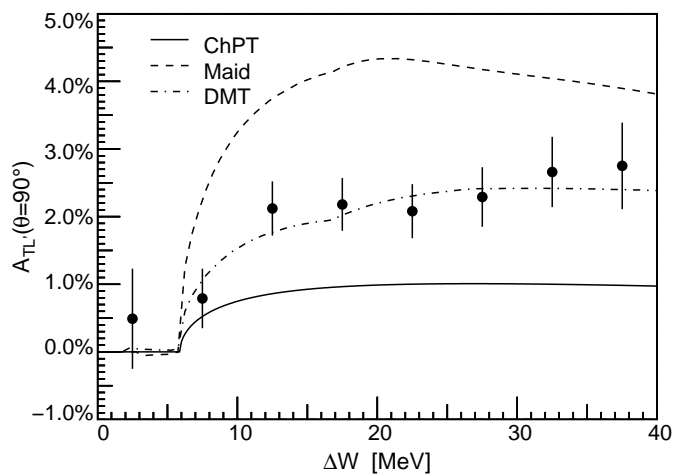}
  \caption{The beam helicity asymmetry $A_{LT'}$ vs. center of mass
    energy.  The solid line shows the calculation in HBChPT
    \protect\cite{BKM96a}, the dashed line shows the phenomenological
    model MAID \cite{DHKT99}, the dash-dotted line shows the DMT
    model\cite{Kamalov:2001qg}.}
  \label{fig:asym}
\end{figure}

Table \ref{tab:asym} and Fig.~\ref{fig:asym} give the results for the
helicity asymmetry $A_{LT'}$. These data are the first experimental
confirmation of the unitary cusp in an observable which is
proportional to the imaginary part of the $s$-wave
amplitudes. Unfortunately, due to the small size of this asymmetry the
energy bins for this variable had to be made larger than for the
unpolarized observables and a detailed determination of the shape is
precluded. The data point below $n\pi^+$ threshold is consistent with
the expected value of zero.  $A_{LT'}$ is clearly underestimated by
the HBChPT calculation \cite{BKM96a}, this is another hint that a
higher order calculation is needed. The MAID model \cite{DHKT99}
overestimates $A_{LT'}$ by nearly a factor of 2. This clearly shows
the necessity to include polarization observables into
phenomenological fits to fix the small multipoles. The DMT model
\cite{Kamalov:2001qg} is in surprising good agreement with the
data. We want to stress, that this is a prediction without knowledge
of the data. In this model, the threshold behavior is generated
dynamically, taking the rescattering effects into account.

In all of the three models $|P_{2}|\gg|P_{5}|$, so that the primary
sensitivity of $A_{LT'}$ is to the imaginary part of the longitudinal
multipole $L_{0+}$.

\section{Summary}

In this experiment, we separated the cross sections $\sigma_0$,
$\sigma_{LT}$, and $\sigma_{TT}$ of neutral pion electroproduction at
$\theta=90^\circ$ at a photon virtuality of $Q^2 =
0.05\,\mathrm{GeV}^2/c^2$ from threshold up to 40\,MeV above
threshold. The cross sections were compared with predictions of Heavy
Baryon Chiral Perturbation Theory \cite{BKM96a} and the
phenomenological models MAID\cite{DHKT99} and DMT
\cite{Kamalov:2001qg}. The discrepancy between HBChPT and the data
suggests the need for higher order calculations.  The phenomenological
models MAID and DMT show reasonable agreement with the separated cross
sections.

Simultaneous, the helicity asymmetry $A_{LT'}(90^\circ)$ was
measured. This observable is sensitive to the imaginary part of the
$s$-wave amplitudes, which was accessed for the first time by a direct
measurement. This asymmetry can be described with impressive agreement
by the DMT model.

\begin{acknowledgement}
\sloppypar This work was supported by the Collaborative Research
Centers 201 and 443 of the Deutsche Forschungsgemeinschaft (DFG) and
by the Federal State of Rhineland-Palatinate, and at MIT in part by
the U.S. Department of Energy under Grant No. DEFG02-94ER40818.
A.\,M.~Bernstein is grateful to the Alexander von Humboldt Foundation
for a Research Award. We would like to thank Ulf-\,G.~Mei\ss{}ner for
discussions and V.\,R.~Brown for helpful comments on the manuscript.
\end{acknowledgement}
\begin{table}
\caption{The separated cross sections $\sigma_0$, $\sigma_{LT}$, and
  $\sigma_{TT}$. Only the model-independent results with cut $75^\circ
  < \theta < 105^\circ$ are given in this table.}
\label{tab:structurefunctions}
\begin{tabular*}{\columnwidth}{@{\extracolsep{\fill}}%
D{.}{.}{1}r@{$~\pm~$}rr@{$~\pm~$}rr@{$~\pm~$}r}
\hline\noalign{\smallskip}
\multicolumn{1}{c}{$\Delta W$} &
\multicolumn{2}{c}{$\sigma_0(90^\circ)$} &
\multicolumn{2}{c}{$\sigma_{LT} (90^\circ)$} &
\multicolumn{2}{c}{$\sigma_{TT}(90^\circ)$} \\
\multicolumn{1}{c}{(MeV)} &
\multicolumn{2}{c}{(nb/sr)} &
\multicolumn{2}{c}{(nb/sr)} &
\multicolumn{2}{c}{(nb/sr)} \\
\noalign{\smallskip}\hline\noalign{\smallskip}
 0.5 &   29.5 &   1.0 &  -13.2 &  0.7 &   -1.4 &   1.5 \\
 1.5 &   43.7 &   0.6 &  -27.3 &  0.4 &   14.8 &   0.9 \\
 2.5 &   60.3 &   2.0 &  -36.9 &  1.4 &   23.6 &   3.1 \\
 3.5 &   88.0 &  14.2 &  -41.7 & 10.0 &   -9.3 &  22.3 \\
 4.5 &  115.2 &   7.7 &  -55.1 &  5.5 &   10.5 &  12.1 \\
 5.5 &  137.7 &  10.4 &  -73.7 &  7.4 &   21.9 &  16.4 \\
 6.5 &  172.9 &   5.6 &  -79.1 &  3.9 &   15.2 &   8.7 \\
 7.5 &  192.9 &  10.6 &  -97.9 &  7.7 &   37.0 &  16.6 \\
 8.5 &  243.9 &   7.1 & -110.6 &  5.1 &   21.4 &  11.2 \\
 9.5 &  288.4 &   7.9 & -124.0 &  5.6 &   34.4 &  12.5 \\
10.5 &  337.0 &  10.9 & -142.2 &  7.7 &   33.4 &  17.1 \\
11.5 &  364.2 &  14.3 & -141.7 & 10.1 &   19.8 &  22.5 \\
12.5 &  407.8 &  26.2 & -151.0 & 18.6 &    8.2 &  41.3 \\
13.5 &  443.9 &  50.5 & -139.7 & 35.8 &  -28.8 &  79.5 \\
14.5 &  464.1 &  25.6 & -159.5 & 20.2 &   40.0 &  44.1 \\
15.5 &  575.5 &  42.6 & -187.4 & 34.3 &  -40.1 &  75.0 \\
16.5 &  618.3 &  38.3 & -208.8 & 31.7 &   26.4 &  67.0 \\
17.5 &  650.4 &  81.7 & -197.2 & 67.6 &   14.3 & 143.1 \\
18.5 &  659.9 &  22.3 & -198.4 & 15.4 &   23.0 &  29.8 \\
19.5 &  738.0 &  54.6 & -205.0 & 35.5 &  -33.8 &  75.8 \\
20.5 &  768.0 &  22.6 & -210.8 & 13.3 &  -37.2 &  29.4 \\
21.5 &  806.0 &  35.5 & -245.5 & 21.1 &    7.4 &  47.2 \\
22.5 &  928.0 &  42.1 & -272.0 & 25.4 &   -1.0 &  57.0 \\
23.5 &  971.1 &  59.2 & -237.6 & 35.4 &  -48.1 &  82.4 \\
24.5 & 1018.3 &  46.2 & -256.4 & 27.6 &    6.1 &  64.3 \\
25.5 & 1113.4 &  79.0 & -250.8 & 48.9 &  -70.6 & 113.6 \\
26.5 & 1190.2 &  61.8 & -318.4 & 38.2 &   24.0 &  88.9 \\
27.5 & 1264.3 &  69.3 & -329.4 & 42.9 &   -3.6 &  99.6 \\
28.5 & 1329.0 &  79.2 & -331.9 & 47.2 &  -16.9 & 113.4 \\
29.5 & 1376.6 &  66.4 & -322.7 & 39.6 &  -62.8 &  95.1 \\
30.5 & 1427.1 &  75.0 & -318.1 & 44.7 & -135.2 & 107.4 \\
31.5 & 1580.4 &  54.0 & -370.3 & 32.6 &  -49.9 &  73.9 \\
32.5 & 1722.7 &  54.7 & -422.6 & 31.4 & -121.1 &  73.8 \\
33.5 & 1879.8 &  71.2 & -409.8 & 41.0 & -114.3 &  96.2 \\
34.5 & 2003.2 &  70.2 & -470.2 & 40.3 & -167.6 &  94.7 \\
35.5 & 2027.6 &  95.3 & -466.7 & 54.8 &  -50.5 & 128.6 \\
36.5 & 2169.0 & 110.7 & -464.3 & 63.7 & -209.0 & 149.4 \\
37.5 & 2146.9 & 115.9 & -464.2 & 66.6 & -167.4 & 156.3 \\
38.5 & 2223.2 & 127.1 & -477.1 & 72.1 &  -51.4 & 168.3 \\
39.5 & 2492.2 & 140.1 & -411.6 & 71.7 & -396.6 & 182.6 \\
\noalign{\smallskip}\hline
\end{tabular*}
\end{table}
\begin{table}
\caption{The helicity asymmetry $A_{LT'}$ at $\theta=90^\circ$.}
\label{tab:asym}
\begin{center}
\begin{tabular}{rr@{$~\pm~$}r}
\hline\noalign{\smallskip}
\multicolumn{1}{c}{$\Delta W$} & \multicolumn{2}{c}{$A_{LT`}$}\\
\multicolumn{1}{c}{(MeV)} & \multicolumn{2}{c}{(\%)}\\
\noalign{\smallskip}\hline\noalign{\smallskip}
 2.5 & 0.49 & 0.74 \\
 7.5 & 0.79 & 0.44 \\
12.5 & 2.12 & 0.40 \\
17.5 & 2.18 & 0.39 \\
22.5 & 2.08 & 0.40 \\
27.5 & 2.29 & 0.44 \\
32.5 & 2.66 & 0.52 \\
37.5 & 2.75 & 0.64 \\
\noalign{\smallskip}\hline
\end{tabular}
\end{center}
\end{table}

\end{document}